\documentclass{article}
 
\usepackage{amsmath,amsthm,amssymb}

\usepackage{cite}

\pdfoutput=1

\usepackage{graphics,graphicx}
\usepackage{epsfig}
\usepackage{multicol}
\usepackage{color}
\makeatletter
\@addtoreset{equation}{section}
\makeatother

\usepackage{braket}
\usepackage{bm}
\usepackage{graphicx}

\usepackage{subcaption}
\usepackage{relsize}

\usepackage{soul}

\usepackage{lineno}

\usepackage{authblk}
\usepackage{hyperref}

\begin{document}

\title{The $(1+3)$-dimensional `quantum principle of relativity’ is Einstein's principle of relativity}
\author[1,2]{Matthew J. Lake\footnote{matthew.lake@ubbcluj.ro}} 
\affil[1]{Faculty of Physics, Babe\c s-Bolyai University, Mihail Kog\u alniceanu Street 1, 400084 Cluj-Napoca, Romania}
\affil[2]{School of Physics, Sun Yat-Sen University, Guangzhou 510275, People's Republic of China}

\maketitle

\begin{abstract}
We show that the $(1+3)$-dimensional `superboost' operators, proposed in Dragan and Ekert's most recent work on superluminal reference frames \cite{Dragan:2022txt}, are simply the canonical Lorentz boosts, expressed in nonstandard notation. 
Their $(1+3)$-dimensional `superflip', which is claimed to interchange time and space dimensions for a superluminal observer, travelling with infinite speed, is equivalent to applying the identity operator together with an arbitrary relabeling. 
Physically, it corresponds to staying put within the canonical rest frame, then renaming space as `time' and time as `space'. 
We conclude that their extension of the `quantum principle of relativity', proposed in earlier work on $(1+1)$-dimensional spacetimes \cite{Dragan:2019grn}, to ordinary Minkowski space \cite{Dragan:2022txt}, is simply Einstein's principle of relativity, proposed in 1905.  
\end{abstract}


\tableofcontents

\section{Recap of Dragan and Ekert's work} \label{Sec.1}

In their most recent work \cite{Dragan:2022txt} on the quantum principle of relativity \cite{Dragan:2008iav,Dragan:2019grn,Dragan:2022}, Dragan and Ekert et al. (DE) consider the following transformation of the canonical rest frame coordinates, $x^{\mu} = [ct,{\bf r}]^{\rm T}$, for $V = |{\bf V}|$ in the range $1 < V/c < \infty$,
\begin{eqnarray} \label{superboost}
\left\{ct,{\bf r}\right\} \rightarrow \left\{
\begin{aligned} 
r' = \frac{Vt - \frac{{\bf r}.{\bf V}}{V}}{\sqrt{\frac{V^2}{c^2}-1}} \, , \\
c{\bf t}' = {\bf r} - \frac{{\bf r}.{\bf V}}{V^2}{\bf V} + \frac{\frac{{\bf r}.{\bf V}}{Vc}-\frac{ct}{V}}{\sqrt{\frac{V^2}{c^2}-1}}{\bf V} \, . 
\end{aligned} \right.
\end{eqnarray} 
They call this a {\it superboost}. 
They then note that the $V \rightarrow \infty$ limit of the superboost (\ref{superboost}) yields
\begin{eqnarray} \label{superflip}
\left\{ct,{\bf r}\right\} \rightarrow \left\{
\begin{aligned} 
r' = ct \, , \\
c{\bf t}' = {\bf r} \, ,
\end{aligned} \right.
\end{eqnarray}
which they refer to as a {\it superflip}. 

Taking the differentials of the quantities $r'$ and $c{\bf t}'$, which are {\it defined} by Eqs. (\ref{superboost}), squaring them, and subtracting one from the other, it is straightforward to verify that the relation
\begin{eqnarray} \label{invariant_line-element-1}
c^2({\rm d}t)^2 - {\rm d}{\bf r}.{\rm d}{\bf r} = -c^2{\rm d}{\bf t}'.{\rm d}{\bf t}' + ({\rm d}r')^2 
\end{eqnarray}
holds, in general, for any superluminal boost speed $V$.

Speaking of the expressions on the left- and right-hand sides of this equation, DE claim that each represents a ``different metric'' \cite{Dragan:2022txt} and, hence, that the spacetime metric seen by a superluminal observer is equivalent to that observed in a subluminal frame, but with the time- and space-like dimensions interchanged. 
Their claim is that this results in a ``signature flip'', 
\begin{eqnarray} \label{sig_flip-1}
\eta_{\mu\nu} = {\rm diag}(1,-1,-1,-1) \rightarrow -\eta'_{\mu\nu} = {\rm diag}(-1,-1,-1,1) \, , 
\end{eqnarray}
in the matrix of metric components, so that the pre- and post-superboost spacetime intervals are related via 
\begin{eqnarray} \label{sig_flip-2} 
\eta_{\mu\nu}x^{\mu}x^{\nu} = -\eta'_{\mu\nu}x'^{\mu}x'^{\nu} \, . 
\end{eqnarray}
(See Eq. (7) in \cite{Dragan:2022txt}.) 
Hence, they claim that the superboost (\ref{superboost}) and in particular its infinite velocity limit, the superlfip (\ref{superflip}), effectively maps a $(1+3)$-dimensional spacetime into a $(3+1)$-dimensional spacetime \cite{Dragan:2008iav,Dragan:2019grn,Dragan:2022,Dragan:2022txt}. 

\section{The equivalence of boosts and `superboosts'} \label{Sec.2}

Unfortunately, Eqs. (\ref{superboost})-(\ref{superflip}) are ambiguous because they include a nonstandard relabeling of the linearly transformed rest frame coordinates, $x'^{\mu} = [ct',{\bf r}']^{\rm T} = B^{\mu}{}_{\nu}({\bf V})x^{\nu}$, where $B^{\mu}{}_{\nu}({\bf V})$ is the transformation matrix corresponding to the boost velocity ${\bf V}$, in addition to the linear transformation itself. 

In standard notation \cite{French:1968}, the application of the linear map $B$ is denoted as
\begin{eqnarray} \label{explicit-1}
x^{\mu} \rightarrow x'^{\mu} = B^{\mu}{}_{\nu}({\bf V})x^{\nu} \, , 
\end{eqnarray}
or, equivalently, as
\begin{eqnarray} \label{explicit-2}
ct \rightarrow ct' = B^{0}{}_{\nu}({\bf V})x^{\nu} \, , \quad {\bf r} \rightarrow {\bf r}' = B^{(1,2,3)}{}_{\nu}({\bf V})x^{\nu} \, , 
\end{eqnarray}
in vector notation. 
DE's introduction of the new symbols, $r'$ and $c{\bf t}'$, represents an additional step in which the linearly transformed variables, $x'^{\mu} = B^{\mu}{}_{\nu}({\bf V})x^{\nu}$, are relabeled according to 
\begin{eqnarray} \label{relabelling-1}
\left\{x'^{\mu}\right\} = \left\{ct',{\bf r}'\right\} \rightarrow \left\{y'^{\mu}\right\} = \left\{r',c{\bf t}'\right\} \, . 
\end{eqnarray}

For absolute clarity, we stress that the individual components of the primed and unprimed $4$-vectors, $x^{\mu} = [ct,{\bf r}]^{\rm T} = [ct,x,y,z]^{\rm T}$ and $x'^{\mu} = [ct',{\bf r}']^{\rm T} = [ct',x',y',z']^{\rm T}$, are related via 
\begin{eqnarray} \label{explicit-2}
ct \rightarrow ct' = B^{0}{}_{\nu}({\bf V})x^{\nu} \, , \, \, \, x \rightarrow x' = B^{1}{}_{\nu}({\bf V})x^{\nu} \, , \, \, \, y \rightarrow y' = B^{2}{}_{\nu}({\bf V})x^{\nu} \, , \, \, \, z \rightarrow z' = B^{3}{}_{\nu}({\bf V})x^{\nu} \, ,
\end{eqnarray}
in the standard notational convention \cite{French:1968}. 
This notation is {\it unambiguous}, as it uniquely matches a primed symbol, $x'^{\mu}$, with a single component of the transformed rest frame $4$-vector, $B^{\mu}{}_{\nu}({\bf V})x^{\nu}$. 

By contrast, it is not possible to determine, from Eqs. (\ref{superboost}) alone, how the former are related, individually, to the new labels $y'^{\mu} = [c{\bf t}',r']^{\rm T}$. 
There exist $4! = 24$ possible identifications between the individual elements of the sets $\left\{x'^{\mu}\right\}$ and $\left\{y'^{\mu}\right\}$, and, therefore, 24 possible interpretations of DE's proposed superboosts. 
Hence, in order to determine the physical content of the superboost equations (\ref{superboost}), we must undo this relabeling and focus, instead, on the underlying linear transformation.  

To achieve this, we note the following. 
First, if $x'^{\mu} = [ct',{\bf r}']^{\rm T}$ denote the linearly transformed rest frame coordinates $B^{\mu}{}_{\nu}({\bf V})x^{\nu}$, in accordance with standard notation in the relativity literature \cite{French:1968}, then the post-superboost line element may be written as
\begin{eqnarray} \label{post-boost-1}
({\rm d}s')^2 = c^2({\rm d}t')^2 - {\rm d}{\bf r}'.{\rm d}{\bf r}' \, . 
\end{eqnarray}
Second, we stress that DE regard the line element seen by the superluminal observer as $-c^2{\rm d}{\bf t}'.{\rm d}{\bf t}' + ({\rm d}r')^2$ (not, for example, $c^2{\rm d}{\bf t}'.{\rm d}{\bf t}' - ({\rm d}r')^2$) \cite{Dragan:2022txt}.
\footnote{Thanks to Andrzej Dragan for personal correspondence, clarifying this point.} 
This allows us to write 
\begin{eqnarray} \label{post-boost-2}
({\rm d}s')^2 = -c^2{\rm d}{\bf t}'.{\rm d}{\bf t}' + ({\rm d}r')^2 \, . 
\end{eqnarray}
Combining Eq. (\ref{invariant_line-element-1}) with (\ref{post-boost-1})-(\ref{post-boost-2}) then gives
\begin{eqnarray} \label{invariant_line-element-2}
({\rm d}s')^2 = c^2({\rm d}t')^2 - {\rm d}{\bf r}'.{\rm d}{\bf r}' = c^2({\rm d}t)^2 - {\rm d}{\bf r}.{\rm d}{\bf r} = ({\rm d}s)^2 \, . 
\end{eqnarray}

Equation (\ref{invariant_line-element-2}) effectively undoes the nonstandard relabeling imposed in \cite{Dragan:2022txt}, which allows us to see the physical content of Eq. (\ref{invariant_line-element-1}) more clearly. 
It is revealed that the line element remains invariant under the application of the superboost (\ref{superboost}), which means that the spacetime metric remains invariant also \cite{Nakahara:2003}. 

But how can this be? 
Transformations that leave the metric invariant are {\it symmetries} of the spacetime. 
The symmetries of $(1+3)$-dimensional Minkowski space are well known and are described by elements of the Poincar{\' e} group \cite{Jones:1998}. 
This is composed of two subgroups, the group of spacetime translations and the Lorentz group, $SO(1,3)$. 
The Lorentz group, in turn, contains both ordinary spatial rotations and Lorentz boosts \cite{Nakahara:2003,Jones:1998}. 
Therefore, if the boosts described by Eqs. (\ref{superboost}) represent symmetries of Minkowski space, then they must be elements of the Lorentz group. 
In particular, they must be elements of the proper isochronous Lorentz group $SO^{+}(1,3)$, as they do not include parity (P) or time-inversion (T) symmetries.

To see that this is indeed the case, let us define the velocity {\it dual} to ${\bf V}$ as
\begin{eqnarray} \label{dual_vel}
{\bf v} =\frac{c^2}{V^2} {\bf V} \, , \quad \left(v^2 = \frac{c^4}{V^2}\right) \, . 
\end{eqnarray}
Since the superluminal velocity satisfies the condition $V/c > 1$, the dual velocity is always subluminal, $v/c < 1$.
The unit 3-vector that defines the direction of a boost, whether superluminal or subluminal, is then  
\begin{eqnarray} \label{n}
{\bf n} = \frac{{\bf V}}{V} = \frac{{\bf v}}{v} \, . 
\end{eqnarray}

Equations (\ref{superboost}) can now be rewritten in terms of ${\bf n}$ and the would-be superluminal boost speed, $V > c$, as
\begin{eqnarray} \label{superboost*}
r' = \frac{Vt - {\bf r}.{\bf n}}{\sqrt{\frac{V^2}{c^2}-1}} \, , \quad
c{\bf t}' = {\bf r} - ({\bf r}.{\bf n}) \, {\bf n} + \frac{\frac{{\bf r}.{\bf n}}{c/V}-ct}{\sqrt{\frac{V^2}{c^2}-1}}{\bf n} \, . 
\end{eqnarray}
Using (\ref{dual_vel}), Eqs. (\ref{superboost*}) become
\begin{eqnarray} \label{superboost**}
r' = \frac{ct - \frac{v}{c}{\bf r}.{\bf n}}{\sqrt{1-\frac{v^2}{c^2}}} \, , \quad
c{\bf t}' = {\bf r} - ({\bf r}.{\bf n}) \, {\bf n} + \frac{{\bf r}.{\bf n}-vt}{\sqrt{1-\frac{v^2}{c^2}}}{\bf n} \, , 
\end{eqnarray}
or, equivalently,
\begin{eqnarray} \label{superboost***}
r' = \frac{ct - \frac{{\bf r}.{\bf v}}{c}}{\sqrt{1-\frac{v^2}{c^2}}} \, , \quad
c{\bf t}' = {\bf r} - \frac{{\bf r}.{\bf v}}{v^2} {\bf v} + \frac{\frac{{\bf r}.{\bf v}}{v^2}-t}{\sqrt{1-\frac{v^2}{c^2}}}{\bf v} \, .
\end{eqnarray}

The expressions on the right-hand sides of these equations are equivalent to those on the right-hand sides of the canonical Lorentz boosts \cite{Dragan:2008iav,Dragan:2019grn,Dragan:2022,Dragan:2022txt}, but what about the symbols on the left? 
These must be identified, in some way, with the linearly transformed rest frame coordinates, $ct' = B^{0}{}_{\nu}({\bf v})x^{\nu}$ and ${\bf r}' = B^{(1,2,3)}{}_{\nu}({\bf v})x^{\nu}$. 
As stated above, there are $4! = 24$ possible identifications between the individual elements of the sets $\left\{x'^{\mu}\right\} = \left\{ct',x',y',z'\right\}$ and $\left\{y'^{\mu}\right\} = \left\{ct'_1,ct'_2,ct'_3,r'\right\}$, but only 6 of these satisfy the requirement 
$c^2({\rm d}t)^2 - {\rm d}{\bf r}.{\rm d}{\bf r} = -c^2{\rm d}{\bf t}'.{\rm d}{\bf t}' + ({\rm d}r')^2$ (\ref{invariant_line-element-1}) \cite{Dragan:2022txt}. 
One of these identifications is 
\begin{eqnarray} \label{identifications-1}
r' = ct' \, , \quad ct'_1 = x' \, , \quad ct'_2 = y' \, , \quad ct'_3 = z' \, , 
\end{eqnarray}
and the other 5 correspond to permutations of the symbols $\left\{ct'_1,ct'_2,ct'_3\right\}$. 
Any one of these 6 possibilities may be written in a condensed form, using vector notation, as 
\begin{eqnarray} \label{identifications-2}
r' = ct' \, , \quad c{\bf t}' = {\bf r}' \, . 
\end{eqnarray}

Note, again, that these relations {\it define} the symbols $r'$ and $c{\bf t}'$. 
The latter have no physical, or mathematical content, and represent only a change in notational convention. 
Comparing them with Eqs. (\ref{superflip}), we obtain 
\begin{eqnarray} \label{superflip*}
ct \rightarrow r' = ct' = ct \, , \quad {\bf r} \rightarrow c{\bf t}' = {\bf r}' = {\bf r} \, ,
\end{eqnarray}
in the $V \rightarrow \infty$ ($v \rightarrow 0$) limit. 
Thus, the actual linear transformation underlying DE's superflip operation (\ref{superflip}), minus the nonstandard relabeling, is simply the identity operator. 
This is not surprising, as it corresponds to the $V \rightarrow \infty$ limit of their superboost operator (\ref{superboost}). 
This is simply the $v \rightarrow 0$ limit of the canonical Lorentz boost. 
For $1 < V/c < \infty$ ($1 > v/c > 0$), substituting from (\ref{identifications-2}) into (\ref{superboost***}) yields the standard expressions
\begin{eqnarray} \label{Lorentz_boost}
ct' = \frac{ct - \frac{{\bf r}.{\bf v}}{c}}{\sqrt{1-\frac{v^2}{c^2}}} \, , \quad
{\bf r}' = {\bf r} - \frac{{\bf r}.{\bf v}}{v^2} {\bf v} + \frac{\frac{{\bf r}.{\bf v}}{v^2}-t}{\sqrt{1-\frac{v^2}{c^2}}}{\bf v} \, .
\end{eqnarray}

\section{The $(1+1)$-dimensional limit} \label{Sec.3}

Interestingly, the criticisms above do not apply to DE's earlier work on superboosts in $(1+1)$-dimensional spacetimes. 
In \cite{Dragan:2019grn}, they proposed the $(1+1)$-dimensional superboost transformations
\begin{eqnarray} \label{superboost_2D}
ct \rightarrow ct' = \pm \frac{V}{|V|} \frac{ct - \frac{V}{c}x}{\sqrt{\frac{V^2}{c^2}-1}} \, , \quad x \rightarrow x' = \pm \frac{V}{|V|} \frac{x - Vt}{\sqrt{\frac{V^2}{c^2}-1}} \, , 
\end{eqnarray}
whose $V \rightarrow \infty$ limit yields
\begin{eqnarray} \label{superflip_2D}
ct \rightarrow ct' = x \, , \quad x \rightarrow x' = ct \, ,
\end{eqnarray}
with an appropriate choice of sign convention.

The transformation equations (\ref{superboost_2D}) are not ambiguous, as it is clear that $x^{0} = ct$ is mapped according to $x^{0} \rightarrow x'^{0} = B^{0}{}_{\nu}x^{\nu}$ and that $x^{1} = x$ is mapped according to $x^{1} \rightarrow x'^{1} = B^{1}{}_{\nu}x^{\nu}$, where the transformation matrix $B^{\mu}{}_{\nu}(V)$ is given by
\begin{eqnarray} \label{B(V)_matrix}
B^{\mu}{}_{\nu}(V) = \pm \frac{V}{|V|}  \frac{1}{{\sqrt{\frac{V^2}{c^2}-1}}}
\left[
\begin{matrix}
1 & -\frac{V}{c} \\
-V & 1
\end{matrix}
\right] \, . 
\end{eqnarray}
Applying the same change of variables as before, by rewriting Eqs. (\ref{superboost_2D}) in terms of the dual velocity, 
\begin{eqnarray} \label{dual_v_2D}
v = \frac{c^2}{V} \, , 
\end{eqnarray}
we obtain 
\begin{eqnarray} \label{superboost_2D*}
ct \rightarrow ct' = \frac{x - vt}{\sqrt{1-\frac{v^2}{c^2}}} \frac{v}{|v|} \, , \quad x \rightarrow x' = \frac{ct - \frac{v}{c}x}{\sqrt{1-\frac{v^2}{c^2}}} \frac{v}{|v|}  \, , 
\end{eqnarray}
where we again take the negative signs in Eqs. (\ref{superboost_2D}), following \cite{Dragan:2019grn}. 

The expressions on the right-hand sides of Eqs. (\ref{superboost_2D*}) are the same as the expressions on the right-hand sides of the canonical $(1+1)$-dimensional Lorentz boosts, multiplied by an extra factor of $v/|v|$. 
However, in this case, the canonical transformation of $ct$, $ct \rightarrow \gamma(v)(ct - \frac{v}{c}x)$, is applied to $x$, and the canonical transformation of $x$, $x \rightarrow \gamma(v)(x - vt)$, is applied to $ct$, for $v > 0$. 
For $v < 0$, the corresponding improper Lorentz boosts are applied instead.

In this case, both the subluminal Lorentz boosts and the superluminal boosts can be meaningfully applied, in different regimes, since the latter are mathematically distinct from the former. 
Either set of transformations can be written in terms of a `subluminal' speed $|v| < c$, or a `superluminal' speed $|V| > c$, but these variables have different physical meanings in each sector. 
For subluminal boosts, the physical velocity with respect to the rest frame is $v = {\rm d}x/{\rm d}t|_{x'=0} < c$ and $V = c^2/v > c$ is the dual variable, whereas, for superluminal boosts, the physical velocity is $V = {\rm d}x/{\rm d}t|_{x'=0} > c$ and $v = c^2/V < c$ is the dual variable. 
In particular, the $V \rightarrow \infty$ ($v \rightarrow 0$) limit of Eq. (\ref{B(V)_matrix}), using DE's sign convention \cite{Dragan:2019grn}, yields the transformation matrix 
\begin{eqnarray} \label{B(V)_matrix*}
B^{\mu}{}_{\nu}(V \rightarrow \infty) = 
\left[
\begin{matrix}
0 & 1 \\
1 & 0
\end{matrix}
\right] \, ,
\end{eqnarray}
instead of the identity matrix, as in the $V \rightarrow \infty$ ($v \rightarrow 0$) limit of their $(1+3)$-dimensional model \cite{Dragan:2022txt}. 
The resulting transformation of the spacetime line element is 
\begin{eqnarray} \label{line-element_transf}
({\rm d}s)^2 = c^2({\rm d}t)^2 - ({\rm d}x)^2 \rightarrow ({\rm d}s')^2 &=& c^2({\rm d}t')^2 - ({\rm d}x')^2 
\nonumber\\
 &=& ({\rm d}x)^2 - c^2({\rm d}t)^2 
 \nonumber\\
 &=& -({\rm d}s)^2 \, , 
\end{eqnarray}
for any value of $1 < V/c < \infty$ (or, equivalently, $1 > v/c > 0$). 
In this model, the local line element is {\it not} conserved under a superluminal velocity boost and the superboost defined by Eqs. (\ref{superboost_2D}) does not represent a symmetry of the spacetime. 
It is therefore not an element of $SO^{+}(1,3)$.

For absolute clarity, we stress that Eqs. (\ref{line-element_transf}) are obtained by applying the standard definition of a linear map \cite{Axler:2015} to the infinitesimal displacement vector, ${\rm d}{\bf x} = {\rm d}x^{\nu}{\bf e}_{\nu}$, where ${\bf e}_{\nu} = \boldsymbol{\partial}/{\boldsymbol{ \partial} x^{\nu}}$ is the vector tangent to the line of constant $x^{\nu}$ \cite{Nakahara:2003}. 
This gives  
\begin{eqnarray} \label{}
B: {\rm d}{\bf x} = {\rm d}x^{\nu}{\bf e}_{\nu}(x) \rightarrow {\rm d}{\bf x}' = {\rm d}x^{\nu}B({\bf e}_{\nu})
\nonumber\\
= {\rm d}x^{\nu} {\bf e}_{\mu} B^{\mu}{}_{\nu}(V)
\nonumber\\
=: {\rm d}x^{\nu} {\bf e}'_{\nu}
\nonumber\\
= B^{\mu}{}_{\nu}(V){\rm d}x^{\nu} {\bf e}_{\mu} 
\nonumber\\
=: {\rm d}x'^{\mu} {\bf e}_{\mu} \, .
\end{eqnarray}
Note that an {\it active transformation} of a vector space $\mathcal{V}$ is represented by an endomorphism on $\mathcal{V}$. 
The endomorphism $B: \mathcal{V} \rightarrow \mathcal{V}$ acts only on the basis vectors of $\mathcal{V}$, and not on their associated components \cite{Axler:2015}. 
The latter are just real or complex numbers, and, therefore, are not elements of the space. 
The condition $B({\rm d}{\bf x}) = B({\rm d}x^{\nu}{\bf e}_{\nu}) = {\rm d}x^{\nu}B({\bf e}_{\nu})$ then follows from the fact that a linear map is a homogeneous map of degree 1 \cite{Axler:2015}. 

In this case, the space $\mathcal{V}$ is two-dimensional Minkowski space and the relevant basis vectors are the tangent vectors, ${\bf e}_{0} = \boldsymbol{\partial}/{\boldsymbol{ \partial} x^{0}}$ and ${\bf e}_{1} = \boldsymbol{\partial}/{\boldsymbol{ \partial} x^{1}}$, which transform according to 
\begin{eqnarray} \label{}
B: {\bf e}_{\nu} \rightarrow B({\bf e}_{\nu} ) = {\bf e}'_{\nu} = {\bf e}_{\mu} B^{\mu}{}_{\nu}(V) \, . 
\end{eqnarray}
Nonetheless, it is conventional, in the relativity literature, to absorb the action of the transformation matrix, $B^{\mu}{}_{\nu}(V)$, into the definition of the transformed coordinates, ${\rm d}x'^{\mu} = B^{\mu}{}_{\nu}(V){\rm d}x^{\nu}$. 
It is essential to recognise, however, that one must not transform {\it both} the basis vectors and the coordinates. 
This is equivalent to applying the linear map $B$ {\it twice}.
Likewise, one must not apply $B$ to the basis vectors and its transpose, $B^{\rm T}$, to the components, or vice versa. 
This generates a {\it passive transformation}, which leaves all physical observables invariant \cite{Axler:2015}. 
For orthogonal matrices, this is simply equivalent to applying the identity transformation: an immediately relevant example is the transformation matrix $B^{\mu}{}_{\nu}(V \rightarrow \infty)$ (\ref{B(V)_matrix*}), for which $B(V \rightarrow \infty)B^{\rm T}(V \rightarrow \infty) = B^{\rm T}(V \rightarrow \infty)B(V \rightarrow \infty) = I$. 

Though explicit details of the calculations performed are not given in \cite{Dragan:2022txt}, the form of Eq. (\ref{sig_flip-2}), which contains both primed coordinates ${\rm d}x'^{\mu}{\rm d}x'^{\nu}$ and a primed matrix of metric components, $-\eta'_{\mu\nu}$, suggests that DE may have defined $({\rm d}s')^2 := {\rm d}{\bf x}^{\rm T}B({\bf V})B^{\rm T}({\bf V})\eta B({\bf V})B^{\rm T}({\bf V}){\rm d}{\bf x}$, as their post-superboost line element, in $(1+3)$ dimensions. 
This is not a valid procedure, for the reasons outlined above (see also \cite{Axler:2015}, or similar literature on linear maps). 
In particular, it yields $({\rm d}s)^2 = {\rm d}{\bf x}^{\rm T}\eta{\rm d}{\bf x} \rightarrow ({\rm d}s')^2 := {\rm d}{\bf x}^{\rm T}B(V \rightarrow \infty)B^{\rm T}(V \rightarrow \infty)\eta B(V \rightarrow \infty)B^{\rm T}(V \rightarrow \infty){\rm d}{\bf x}$, in the infinite velocity limit. 
This is not a meaningful transformation, for {\it any} of the $4! = 24$ possible definitions of $B^{\mu}{}_{\nu}({\bf V})$ that are compatible with Eqs. (\ref{superboost}) -- not only the 6 that correspond to relabelings of the Lorentz transformations, considered in Sec. \ref{Sec.2} -- since the relation $B(V \rightarrow \infty)B^{\rm T}(V \rightarrow \infty) = B^{\rm T}(V \rightarrow \infty)B(V \rightarrow \infty) = I$ {\it always} holds, in all 24 scenarios. 

Though we can only speculate, in the absence of explicit calculations showing their methods, it is possible that the authors' {\it implicit definition} of the post-superboost line element, $({\rm d}s)^2 = {\rm d}{\bf x}^{\rm T}\eta{\rm d}{\bf x} \rightarrow ({\rm d}s')^2 := -{\rm d}{\bf x}'^{\rm T} \eta' {\rm d}{\bf x}' := {\rm d}{\bf x}^{\rm T}B(V \rightarrow \infty)B^{\rm T}(V \rightarrow \infty)\eta B(V \rightarrow \infty)B^{\rm T}(V \rightarrow \infty){\rm d}{\bf x} = {\rm d}{\bf x}^{\rm T} \, I \, \eta \, I \, {\rm d}{\bf x} = ({\rm d}s)^2$ (\ref{sig_flip-2}) \cite{Dragan:2022txt}, misled them into believing that the metric of the spacetime could be altered {\it without} altering the local line element at any point within it. 
However, this is not the case, as there exists a one-to-one correspondence between the metric, $g(\mathfrak{p}) = \braket{{\bf e}_{\mu}(\mathfrak{p}),{\bf e}_{\nu}(\mathfrak{p})}{\rm d}x^{\mu}(\mathfrak{p}) \otimes {\rm d}x^{\nu}(\mathfrak{p})$, at a point in the spacetime manifold $\mathfrak{p} \in \mathcal{M}$, and the local line element obtained from it, $({\rm d}s)^2 = \braket{{\bf e}_{\mu}(\mathfrak{p}),{\bf e}_{\nu}(\mathfrak{p})}{\rm d}x^{\mu}(\mathfrak{p}){\rm d}x^{\nu}(\mathfrak{p})$ \cite{Nakahara:2003}. 

Hence, in any number of spacetime dimensions, the superboost transformation $B$, enacted by the matrix $B^{\mu}{}_{\nu}({\bf V})$, transforms the canonical spacetime line element according to 
\begin{eqnarray} \label{explanation}
&&({\rm d}s)^2 = \braket{{\rm d}{\bf x},{\rm d}{\bf x}} = \braket{{\bf e}_{\mu},{\bf e}_{\nu}}{\rm d}x^{\mu} {\rm d}x^{\nu} = \eta_{\mu\nu}{\rm d}x^{\mu} {\rm d}x^{\nu} 
\nonumber\\
&\rightarrow&({\rm d}s')^2 = \braket{B({\rm d}{\bf x}),B({\rm d}{\bf x})} = \braket{B({\bf e}_{\mu}),B({\bf e}_{\nu})}{\rm d}x^{\mu} {\rm d}x^{\nu}
\nonumber\\
&=& \braket{{\bf e}_{\rho}B^{\rho}{}_{\mu}({\bf V}),{\bf e}_{\sigma}B^{\sigma}{}_{\nu}({\bf V})}{\rm d}x^{\mu} {\rm d}x^{\nu} = \braket{{\bf e}_{\rho},{\bf e}_{\sigma}} B^{\rho}{}_{\mu}({\bf V}){\rm d}x^{\mu} B^{\sigma}{}_{\nu}({\bf V}){\rm d}x^{\nu}
\nonumber\\
&=& \braket{{\bf e}_{\rho},{\bf e}_{\sigma}} {\rm d}x'^{\rho}{\rm d}x'^{\sigma} = \eta_{\rho\sigma}{\rm d}x'^{\rho}{\rm d}x'^{\sigma} \, . 
\end{eqnarray}

This argument is not changed by the fact that the ${\rm d}x^{\mu}$ may be viewed as elements of the cotangent space. 
The latter is isomorphic to the tangent space, and, hence, to $(1+1)$-dimensional Minkowski space. 
Despite these isomorphisms, the cotangent space is the space of 1-forms (not vectors) and, since there is no canonically defined inner product between 1-forms, the latter act as numerical constants with respect to the operation $\braket{ \, . \, , \, . \, }$ \cite{Nakahara:2003}, as well as with respect to the endomorphism on $\mathcal{V}$, $B: \mathcal{V} \rightarrow \mathcal{V}$ \cite{Axler:2015}. 
There is, therefore, no justification at all for acting on them with $(B^{\mu}{}_{\nu}({\bf V}))^{\rm T}$, in the transformation of the line element, as DE appear to have done in \cite{Dragan:2022txt}. 

It is easy to see how doing so in $(1+1)$ dimensions may lead one to believe, erroneously, that the line element $({\rm d}s)^2 = c^2({\rm d}t)^2 - ({\rm d}x)^2$ remains invariant under the application of the superflip (\ref{superflip_2D}), while the metric signature changes. 
In the infinite velocity limit, the superflip matrix $B^{\mu}{}_{\nu}(V \rightarrow \infty)$ (\ref{B(V)_matrix*}) gives $c^2({\rm d}t)^2 \rightarrow ({\rm d}x)^2$ and $({\rm d}x)^2 \rightarrow c^2({\rm d}t)^2$, when applied to the coordinates ${\rm d}x^{\nu}$, and $B^{\rm T}(V \rightarrow \infty){\rm diag}(1,-1)B(V \rightarrow \infty) = {\rm diag}(-1,1)$, when applied to the matrix of metric components, $\eta_{\mu\nu} = {\rm diag}(1,-1)$. 
These two effects cancel one another out, rendering the Minkowski metric $\eta = \eta_{\mu\nu} {\rm d}x^{\mu} \otimes {\rm d}x^{\nu}$ invariant, but give the superficial impression of a metric signature change, $(+-) \rightarrow (-+)$. 
In reality, one has flipped the ordering of the positive and negative eigenvalues in the original matrix $\eta_{\mu\nu}$, but also the ordering of components in the 4-vectors ${\rm d}x^{\mu}$ and ${\rm d}x^{\nu}$. 
This results in a change of ordering convention, but not a change of metric, since the metric is not synonymous with the matrix of its components \cite{Nakahara:2003}. 

The correct transformation is obtained by taking the limit $V \rightarrow \infty$, adopting the sign convention in \cite{Dragan:2019grn} and using $B^{\mu}{}_{\nu}(V \rightarrow \infty)$ (\ref{B(V)_matrix*}) in Eqs. (\ref{explanation}). 
This yields $({\rm d}s)^2 \rightarrow -({\rm d}s)^2$ (\ref{line-element_transf}), as claimed. 
The time-like and space-like vectors are effectively interchanged, but such a transformation does not, and {\it cannot}, leave the spacetime interval invariant.

This helps to explain why the $(1+1)$-dimensional results, presented in \cite{Dragan:2019grn}, do not generalise to $(1+3)$-dimensional Minkowski space, in the way that DE claim in \cite{Dragan:2022txt}. 
In $(1+3)$ dimensions, there are $4! = 24$ possible generalisations of the $(1+1)$-dimensional superflip matrix, $B^{\mu}{}_{\nu}(V \rightarrow \infty)$ (\ref{B(V)_matrix*}), namely
\begin{eqnarray} \label{superflip_matrix*}
B^{\mu}{}_{\nu}(V \rightarrow \infty) = 
\left[
\begin{matrix}
0 & 0 & 0 & 1 \\
0 & 0 & 1 & 0 \\
0 & 1 & 0 & 0 \\
1 & 0 & 0 & 0 
\end{matrix}
\right]
\, , \, 
\left[
\begin{matrix}
0 & 0 & 0 & 1 \\
0 & 1 & 0 & 0 \\
0 & 0 & 1 & 0 \\
1 & 0 & 0 & 0 
\end{matrix}
\right]
\, \dots 
\left[
\begin{matrix}
1 & 0 & 0 & 0 \\
0 & 0 & 1 & 0 \\
0 & 1 & 0 & 0 \\
0 & 0 & 0 & 1 
\end{matrix}
\right]
\, , \, 
\left[
\begin{matrix}
1 & 0 & 0 & 0 \\
0 & 1 & 0 & 0 \\
0 & 0 & 1 & 0 \\
0 & 0 & 0 & 1 
\end{matrix}
\right]
 \, . 
\end{eqnarray}
The final 6 of these preserve the line element, and hence the spacetime metric \cite{Nakahara:2003}, and are equivalent to canonical special relativity under permutations of the symbols denoting the space-like coordinates $\left\{x',y',z'\right\}$ (or relabelings thereof, such as DE's $\left\{ct'_1,ct'_2,ct'_3\right\}$). 
The remaining 18 result in a `superflip' that genuinely changes the spacetime metric. 
These correspond, in fact, to the remaining 18 possible interpretations of Eqs. (\ref{superboost}). 

For example, taking the first matrix in Eqs. (\ref{superflip_matrix*}) as our definition of the superflip operator, $B^{\mu}{}_{\nu}(V \rightarrow \infty)$, we obtain the following transformation of the $(1+3)$-dimensional line element,
\begin{eqnarray} \label{}
({\rm d}s)^2 = c^2({\rm d}t)^2 - ({\rm d}x)^2 - ({\rm d}y)^2 - ({\rm d}z)^2 \rightarrow ({\rm d}s')^2 = c^2({\rm d}t')^2 - ({\rm d}x')^2 - ({\rm d}y')^2 - ({\rm d}z')^2
\nonumber\\
= ({\rm d}z)^2 - ({\rm d}y)^2 - ({\rm d}x)^2 - c^2({\rm d}t)^2 
\nonumber\\
\neq ({\rm d}s)^2 \neq -({\rm d}s)^2\, , 
\end{eqnarray}
which, following the logic in \cite{Dragan:2022txt}, could be relabeled as 
\begin{eqnarray} \label{}
({\rm d}s)^2 = c^2({\rm d}t)^2 - ({\rm d}x)^2 - ({\rm d}y)^2 - ({\rm d}z)^2 \rightarrow ({\rm d}s')^2 = - ({\rm d}x')^2 - ({\rm d}y')^2 - ({\rm d}z')^2 + c^2({\rm d}t')^2 
\nonumber\\
= - ({\rm d}y)^2 - ({\rm d}x)^2 -c^2({\rm d}t)^2 + ({\rm d}z)^2
\nonumber\\
= - c^2({\rm d}t'_3)^2 - c^2({\rm d}t'_2)^2 - c^2({\rm d}t'_1)^2 + ({\rm d}r')^2 \, . 
\end{eqnarray}
Unfortunately, this transformation is still deeply problematic. 
It predicts that the spatial dimension corresponding to the designated `$z$-axis' {\it always} undergoes the superflip, effectively exchanging roles with the rest frame time dimension, regardless of the direction, ${\bf n}$, in which the superboost (\ref{superboost}) is applied. 
Since `$z$' is just a label, this leads to coordinate-dependent predictions, which must be considered as unphysical. 
Similar remarks also apply to the other 17 possible choices of metric-altering superflip (\ref{superflip_matrix*}).

In $(1+1)$ dimensions, however, this problem does not occur due to the unique symmetry of the spacetime: there is only one spatial dimension capable of `flipping' with the temporal dimension. 
In addition, $(1+1)$-dimensional spacetime is the only geometry in which superflipping the temporal dimension with a single dimension of space can result in an overall sign change in the square of the line element, $({\rm d}s)^2 \rightarrow ({\rm d}s')^2 = -({\rm d}s)^2$. 

In summary, we see that, using {\it only} the expressions on the right-hand sides of the canonical Lorentz transformations (\ref{Lorentz_boost}), in the construction of superluminal boost operators -- as in the definitions of DE's superboosts (\ref{superboost}) -- the only sensible generalisations of the matrix $B^{\mu}{}_{\nu}(V \rightarrow \infty)$ (\ref{B(V)_matrix*}) are the 6 possibilities that leave the $(1+3)$-dimensional line element invariant, i.e., those corresponding to the 6 possible relabelings of the {\it canonical} Lorentz boosts, discussed in Sec. \ref{Sec.2}. 
The solution to this problem, we have suggested elsewhere \cite{Lake:2024}, is to look beyond reorderings and relabelings of the canonical transformations, in order to construct a map between time-like and space-like hypersurfaces. 
The former define the paths of objects moving subluminally, with respect to a given observer, while the latter define the paths of objects in superluminal motion.
In $(1+1)$ dimensions both surfaces are simple hyperbolae, but, in $(1+3)$ dimensions, they correspond instead to hyperbolic sections of $1$-sheet and $2$-sheet hyperboloids, respectively. 
By defining a map between $2$-sheet hyperboloids within the light cone and $1$-sheet hyperboloids without, it is possible to obtain a covariant definition of a superluminal boost operator, in ordinary Minkowski space. 
The $(1+1)$-dimensional limit of these transformations yield expressions similar to Eqs. (\ref{superboost_2D}), but without the convention-dependent ambiguities associated with the factors of $\pm V/|V|$ \cite{Lake:2024}. 

Finally, it is worth noting that it is impossible to flip the metric signature of ordinary Minkowski space, in the way that DE claim, because there are no real transformations capable of achieving this. 
A formal proof of this statement appeared in textbooks on linear algebra as early as 1932, but was well known in the literature before that \cite{Turnbull:1932}. 
It was also well known to later generations of researchers, being referenced in the comprehensive review on superluminal transformations by Marchildon et al, in 1983 \cite{Marchildon:1983}. 
In fact, it is straightforward to see that only transformations of the form ${\rm d}x^{\mu} \rightarrow {\rm d}x'^{\mu} = \pm i\Lambda^{\mu}{}_{\nu}{\rm d}x^{\nu}$, where $\Lambda^{\mu}{}_{\nu}$ is a canonical Lorentz transformation, are capable of flipping the $(1+3)$-dimensional signature to the $(3+1)$-dimensional signature. 
To make it appear otherwise, at least superficially, one must define the signature-flipped matrix of metric components, $\eta_{\mu\nu}' = -\eta_{\mu\nu}$ (\ref{sig_flip-1}), as DE did in \cite{Dragan:2022txt}, and then replace $\eta_{\mu\nu}$ with $-\eta_{\mu\nu}'$ in all the relevant equations (see, for example, (\ref{sig_flip-2})). 
Clearly, this changes nothing and is equivalent to yet another relabelling of variables, that is without physical or mathematical content.

Applying the transformation ${\rm d}x^{\mu} \rightarrow {\rm d}x'^{\mu} = \pm i\Lambda^{\mu}{}_{\nu}{\rm d}x^{\nu}$ to a relativistic wave equation, for example, the Klein-Gordon or Dirac equation, is equivalent to applying the transformation $\partial_{\mu} \rightarrow \partial'_{\mu} = \pm i\partial_{\mu}$, while keeping $m$ fixed. 
This, in turn, is equivalent to keeping the metric fixed and mapping $m \rightarrow \pm im$, which is the standard way to define a tachyonic state \cite{Feinberg:1967}. 
In summary, the metric signature may be effectively switched, from $(1+3)$ to $(3+1)$ dimensions, using a pure-imaginary transformation that is equivalent to doing special relativity with pure-imaginary coordinates, but it cannot be flipped using DE's formulae \cite{Dragan:2022txt}. 
This is because the latter are simply the canonical Lorentz transformations, expressed in nonstandard notation. 

\section{Conclusions} \label{Sec.4}

Our conclusions are as follows:

\begin{enumerate}

\item DE's superflip equations, $r' = ct$, $c{\bf t}' = {\bf r}$ (\ref{superflip}), are a relabeling of the tautological expressions $ct = ct$, ${\bf r} = {\bf r}$. 
They are logically equivalent to renaming time as `space' and space as `time' but possess no physical or mathematical content: they represent only a change in notational convention.

\item As already noted in \cite{Dragan:2022txt}, a general superboost with velocity ${\bf V} = V{\bf n}$ ($1 < V/c < \infty$, $|{\bf n}| = 1$), defined therein, is equivalent to a superflip, plus an ordinary Lorentz boost by the dual velocity ${\bf v} =  (c^2/V){\bf n}$.

\item The combination of statements 1 and 2 implies that a general superboost by ${\bf V}$ ($V/c > 1$) is equivalent to a canonical Lorentz boost by ${\bf v}$ ($v/c < 1$). 

\item A canonical Lorentz boost by ${\bf v}$ ($v/c < 1$) can always be rewritten in terms of its associated dual velocity ${\bf V}$ ($V/c > 1$).

\item When this rewriting is performed, we obtain exactly the expressions on the right-hand sides of DE's general superboost equations (\ref{superboost}). 

\item We conclude that these equations constitute a simple relabeling of the canonical subluminal Lorentz boosts, (\ref{Lorentz_boost}). 

\item This conclusion is confirmed by the fact that the spacetime line element remains unchanged after their application, $({\rm d}s')^2 = -c^2{\rm d}{\bf t}'.{\rm d}{\bf t}' + ({\rm d}r')^2 = c^2({\rm d}t)^2 - {\rm d}{\bf r}.{\rm d}{\bf r} = ({\rm d}s)^2$ \cite{Dragan:2022txt}. 

\item The $(1+3)$-dimensional `quantum principle of relativity', proposed in \cite{Dragan:2022txt}, is therefore equivalent to the canonical principle of relativity, proposed by Einstein in 1905. 

\item Interestingly, the criticisms above to not apply to DE's earlier superboost theory, formulated in $(1+1)$ dimensions \cite{Dragan:2019grn}. 
This means that their $(1+3)$-dimensional model is not a valid generalisation of their earlier work, as it does not recover their previous model in the appropriate limit. 

\item The confusion, which led the authors of \cite{Dragan:2022txt} to claim that their $(1+3)$-dimensional theory is a generalisation of the $(1+1)$-dimensional superboost model \cite{Dragan:2019grn}, appears to have been caused by the same relabeling that obscured its equivalence to special relativity. 
A misunderstanding of the nature of {\it active coordinate transformations} \cite{Axler:2015}, as opposed to passive coordinate transformations, also seems to have contributed.

\end{enumerate}

\section*{Acknowledgments} \label{Sec.5}
This work was supported by the Grant of Scientific and Technological Projection of Guangdong Province (China), no. 2021A1515010036.


\end{document}